\documentclass{optica-article}

\journal{opticajournal} 

\articletype{Research Article}

\usepackage{lineno}

\begin{document}

\title{Tunable dynamical tissue phantom for laser speckle imaging}

\author{Soumyajit Sarkar,\authormark{1,*} K. Murali,\authormark{1,*} and Hari M. Varma\authormark{1,$\#$}}

\address{\authormark{1}Dept of Bio-sciences and Bio-engineering, Indian Institute Technology - Bombay, Mumbai, India - 400076\\
\authormark{*} Equal contribution}
\email{\authormark{$\#$}harivarma@iitb.ac.in} 


\begin{abstract*} 

We introduce a novel method to design and implement a tunable dynamical tissue phantom for laser speckle-based \textit{in-vivo} blood flow imaging. This approach relies on Stochastic Differential Equations (SDE) to control a piezoelectric actuator which, upon illuminated with a laser source, generates speckles of pre-defined probability density function and auto-correlation. The validation experiments show that the phantom can generate dynamic speckles that closely replicate both surfaces as well as deep tissue blood flow for a reasonably wide range and accuracy.
\end{abstract*}

\section{Introduction}

Blood flow plays a crucial role in the human body, serving as a vital conduit for delivering oxygen to tissues and facilitating the removal of metabolic waste. It acts as a key bio-marker in medical diagnostics and research. To measure blood flow non-invasively, methods such as Laser Doppler Flowmetry (LDF) \cite{rajan2009review,binzoni2012blood}, Laser Speckle Contrast Imaging (LSCI) \cite{briers1982retinal,parthasarathy2008robust} and Diffuse Correlation Spectroscopy (DCS) \cite{boas1995scattering,durduran2010diffuse,dong2012simultaneously} are widely used. LSCI and its variants including multi-exposure speckle imaging (MESI) use uniform illumination in the region of interest and single-exposure or multi-exposure speckle contrast to measure surface \textit{in-vivo} blood flow changes (up to about 1 mm deep). On the other hand, DCS and its variants \cite{sdobnov2023advances,james2023diffuse,seong2022blood}, such as Speckle Contrast Optical Spectroscopy (SCOS)  \cite{valdes2014speckle, varma2014speckle, kim2023measuring} and multi speckle DCS \cite{murali2020multi, wayne2023massively, murali2020equivalence} are employed to measure deep tissue blood flow. Additionally, interferometric detection methods are also employed to boost the low light signal at the detector end \cite{zhou2021multi,xu2020interferometric,huang2023interferometric, zhao2023interferometric}.

All these modalities require tissue-mimicking phantoms for optimizing and calibration purposes to ensure consistent and comparable results across various systems and over time. 
Currently, liquid phantoms \cite{choe2005diffuse,mcgarry2020tissue,cortese2018liquid} made using a combination of Intralipid, glycerol, and water are routinely used to replicate blood flow. In addition, to replicate a directed flow, the liquid phantoms are flown using a pump.  However, they face challenges like short shelf life, difficulty in replicating arbitrary geometry, and risk of contamination. On the other hand, solid phantoms can last longer, have consistent optical properties, and are easier to handle \cite{sekar2019solid,goldfain2022polydimethylsiloxane}. However, they fail to replicate the dynamic properties of blood flow.   In this context, we propose a new phantom to replicate the dynamic properties of tissues, especially blood flow. Other optical properties of the phantom will be ensured by the nature of the solid PDMS-based phantom itself. We validate the proposed phantom by mimicking the blood flow changes by the phantom for both surface and deep tissue blood flow changes associated with MESI and DCS systems. 

Recently, we introduced a laser speckle simulation method based on the Stochastic Differential Equation (SDE) \cite{murali2022laser} to obtain laser speckles with a predetermined Probability Density Function (PDF) and auto-correlation ($g_1$). In the current study, we extend this methodology to develop a tunable dynamic phantom, capable of simulating various tissue blood flow scenarios. The core idea is to use the SDE-derived signals to actuate a piezo-actuator device to which a diffuser is attached. When a laser source illuminates this phantom, it produces dynamic speckles. By adjusting the parameters of the SDE, we can fine-tune the speckle patterns to represent different blood flow conditions. Depending on the laser source (uniform or point source) and the diffuser employed, our phantom can replicate either superficial or deep tissue blood flow.

\section{Methods}

\subsection{Construction of proposed tunable phantom}

\begin{figure}[ht]
\centering
\includegraphics[height=5cm,width=12cm]{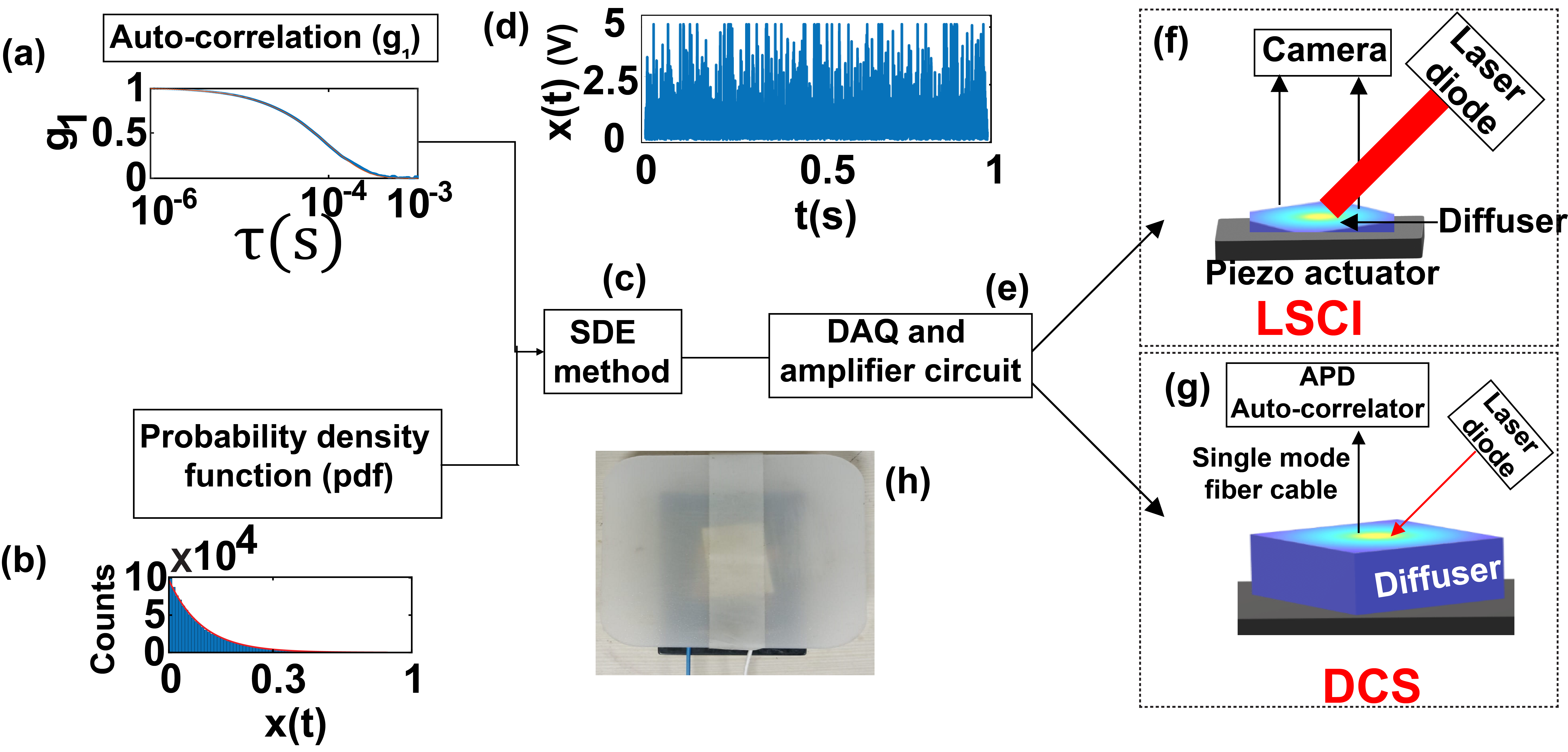}
\caption{Schematic diagram of the tunable dynamical phantom.}
\label{chk1}
\end{figure}

Fig. \ref{chk1} illustrates the schematic of the tunable dynamical tissue phantom. The signal $(x(t))$ as shown in Fig. \ref{chk1}d, is generated using a Stochastic Differential Equation (SDE), as outlined in Ref~\cite{murali2022laser}, with a predetermined auto-correlation (Fig. \ref{chk1}(a)) and PDF (Fig. \ref{chk1}(b)). Briefly, a general SDE is given by \cite{braumann2019introduction}
\begin{equation}
    dx(t) = a(x(t),t)dt + b(x(t),t)dW(t)
    \label{eq1}
\end{equation}
with initial condition $x(0) = x_0$. As given in our earlier work in Ref~\cite{murali2022laser}, we denote  $a$ and $b$ as the drift and diffusion coefficients, respectively. $W$ represents the Wiener process (or Brownian motion) with its increments defined as $dW(t)\sim\sqrt{dt}N(0,1)$, where $N$ is a Gaussian random variable with zero mean and unit variance. The parameters $a=-\alpha(x-\mu)$ and $b =\sqrt{2\mu\alpha x}$ are deduced from the Fokker-Planck equation to ensure that $x(t)$ obeys exponential distribution and exponential auto-correlation. $\alpha$ is the inverse of the characteristic decay time $\tau_{SDE}$ which can be related to blood flow velocity \cite{parthasarathy2008robust} and $\mu$ is mean of $x(t)$.

The signal $x(t)$, obtained by solving SDE in Eq.(\ref{eq1}), is utilized to drive a piezoelectric actuator. This actuator vibrates in response to the signal, thereby replicating the behavior of dynamic scatterers within the tissue. A diffuser (made using PDMS, $TiO_{2}$ and Indian Ink), tailored to replicate the static optical properties of tissue, is positioned on top of the piezoelectric actuator. When a laser illuminates this assembly, it generates dynamic speckles. These speckles are then measured using either LSCI or DCS systems, to characterize the proposed phantom.

\subsection{Signal generation and Actuation}

We solved SDE numerically using the Milstein scheme in Matlab \textsuperscript{\textregistered} for different $\tau_{SDE}$ and multiple realizations \cite{murali2022laser}. These simulated stochastic time series $x(t)$ (shown in  Fig. \ref{chk1}c) are then converted to analog voltages using a data acquisition card (DAQ). The signal is then amplified using op-amp based voltage amplifier (Fig. \ref{chk1}e) before feeding to the Piezoelectric actuator (Piezo-Hap 1919, shown in Fig. \ref{chk1}h). A thin (Fig. \ref{chk1}f) or thick diffuser(Fig. \ref{chk1}g) is attached to the piezo-actuator to replicate the superficial or deep tissue blood flow respectively. Thick diffuser is made of Polydimethylsiloxane(PDMS) \cite{paul2022simple, choe2005diffuse} with optical absorption and reduced scattering coefficients of  $\mu_a=0.003 cm^{-1}$ and $\mu_s'=10 cm^{-1}$ respectively.

\begin{figure}[ht]
\centering
\includegraphics[height=5cm,width=11cm]{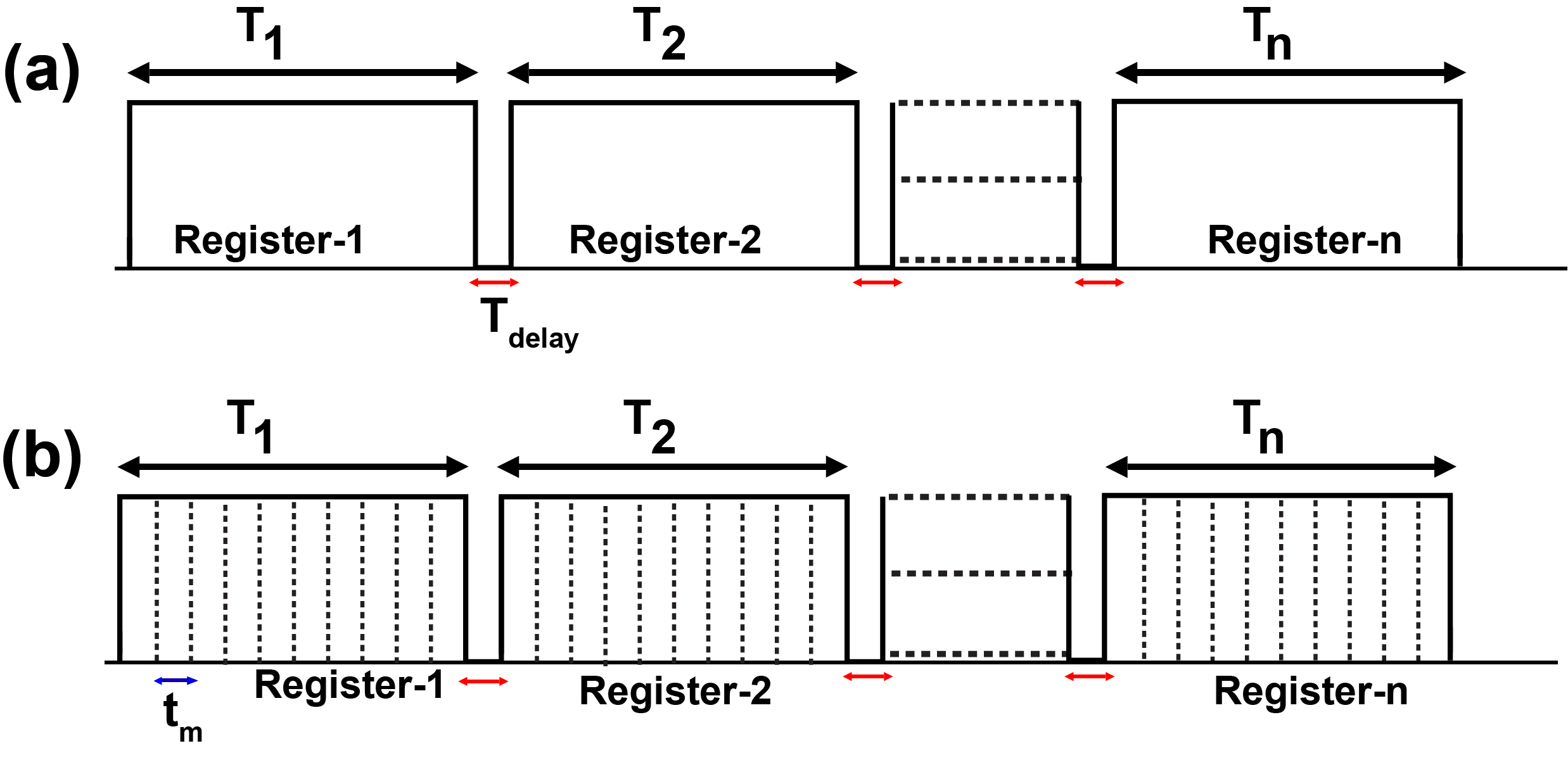}
\caption{Timing diagram of Method-1 and Method-2 utilised to operate the actuator.}
\label{fig:7}
\end{figure}

In the experiments, we employed two methods, as shown in Fig. \ref{fig:7}, to optimize the operation of the actuator. We used a Tektronix AFG1022 generator as a DAQ (Data Acquisition) device to convert the signal $x(t)$ into voltage $v(t)$ for driving the actuator. The DAQ device has 255 registers, denoted by Register-n (in Fig. \ref{fig:7}), each storing $N_{s}=8000$ data points pre-loaded for independent realizations of $x(t)$. In Method-1, as shown in Fig. \ref{fig:7}a, $N_{s}$ values are stored within a time interval of $T_n = 1s$, resulting in a minimum time interval of $dt=0.125ms$. This is followed by a time delay of $T_{delay}=0.1s$ before the next register begins operation. In Method-2, $N_{s}$ values from a register are stored at regular intervals of $t_m=0.1s$ for ten cycles, resulting in $T_n=1s$ with a finer time resolution of $dt=0.0125 ms$. Method-2, as shown in Fig. \ref{fig:7}b, enhances the range of flow speeds that can be realized by the phantom. Note that in Method-2, the same values are repeated over multiple cycles.  

\subsection{Experimental Setup - Surface blood flow}
We used a diode laser at 785 nm (Holmarc Optomechatronics Pvt Limited, India), to provide uniform illumination over a region of $2 \times 2$ cm, and employed a CCD-based camera (Basler acA640-120$\mu m$) with a 50 mm objective lens to measure the scattered intensity. Multi-exposure data was captured at 15 exposure times ranging from 100 $\mu s$ to 2512 $\mu s$, distributed in a log scale. A small region of interest (ROI) of $1\times 1$ mm was mapped to a (100 × 100) pixel area of the camera sensor, with appropriate $f/\#$ to match the speckle-to-pixel ratio. 100 images were acquired at each exposure time to calculate the multi-exposure speckle contrast data for various $\tau_{SDE}$ values. The speckle contrast ($\kappa$) was calculated for each pixel over frames and its mean was computed. Three sets of such experiments were performed for individual $\tau_{SDE}$ values to measure the repeatability. The mean of the trials was fitted to multi-exposure speckle contrast model \cite{parthasarathy2008robust,murali2019recovery} to determine the de-correlation time ($\tau_{c_{\kappa}}$). This process was repeated for multiple flows ($\tau_{SDE}$) to determine the calibration characteristics of the phantom. To show the feasibility of working of the proposed method for \textit{in-vivo} tissue, we have performed a human forearm cuff experiment, wherein the blood flow in human arm was occluded by using a standard sphygmomanometer. This work was undertaken with approval of the Institute Ethical Committee of the Indian Institute of Technology - Bombay, Approval Number: IITB-IEC/2018/017. A single exposure speckle contrast data (measured at 300 $\mu s$ exposure time) was acquired over a period of 180 s over pre-occlusion, occlusion and post-occlusion phases. 

\subsection{Experimental Setup - Deep tissue blood flow}
For DCS measurements, we focused the aforementioned laser source to form a 1 mm diameter point source. We employed a single photon avalanche diode (SPAD) detector (SPCM-AQRH-14-FC) in conjunction with a hardware auto-correlator (correlator.com),  at a source-detector distance of 2 cm. The normalized intensity auto-correlation $g_2(\tau)$ was measured for 30 seconds in each trial. The mean values from three independent trials conducted for various $\tau_{SDE}$ were fitted to the semi-infinite solution to Correlation Diffusion Equation (CDE)\cite{durduran2010diffuse}.

\section{Results}
\subsection{Mimicking Surface blood flow}
\begin{figure}[ht]
\centering
\includegraphics[height=6cm, width=12cm]{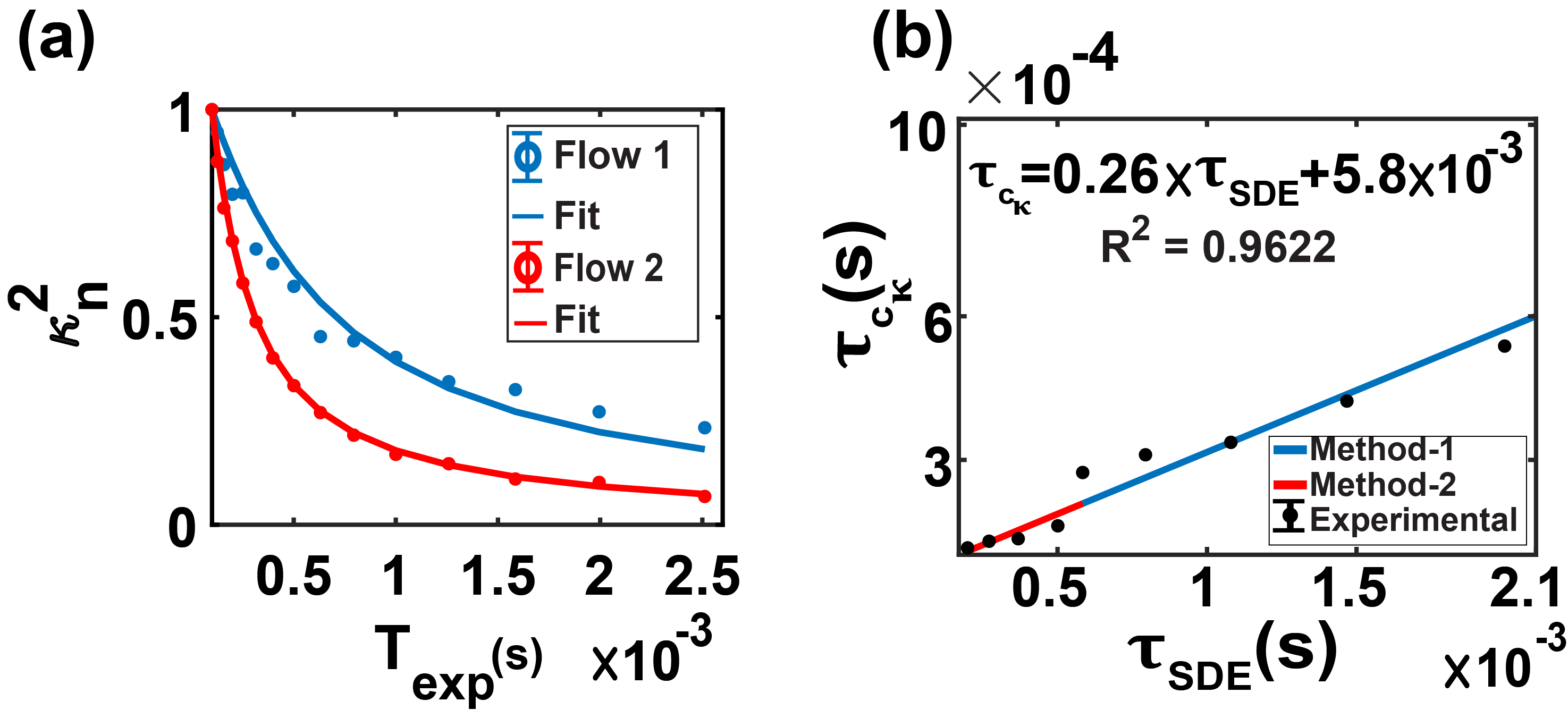}
\caption{Replicating superficial blood flow: (a) Normalized multi-exposure Speckle contrast data($\kappa_n$), along with the fit to determine the de-correlation time $\tau_{c_{\kappa}}$, for two different flow realizations ($\tau_{SDE}$) obtained from the proposed dynamic phantom. (b) Calibration curve of the phantom showing characteristic decay time ($\tau_{SDE}$) used in the SDE model versus the experimentally obtained decay time ($\tau_{c_{\kappa}}$). The high correlation coefficient between them, along with minimal variance (less than $5\%$ relative change with the mean), was observed over three trials.}
\label{fig:3}
\end{figure}

In this section, we present the result of dynamic speckles produced to mimic the surface blood flow. The SDE was solved to obtain $x(t)$ for each flow ($\tau_{SDE}$), which was used to drive the piezo actuator. Upon illuminating with a uniform laser source, the scattered intensities at multiple exposures were captured for 3 trials and speckle contrast$(\kappa)$ was calculated as shown in Fig. \ref{fig:3}(a). It was along with the fit for analytical solution \cite{parthasarathy2008robust} related to surface blood flow changes. The two representative $\tau_{SDE}$ of $1.47 ms$ and $0.1995 ms$, indicated as flow 1 and flow 2 respectively. The fitted decorrelation time ($\tau_{c_{\kappa}})$ for flow 1 and  flow 2 were $0.58 \pm 0.0175 ms$ and $0.268 \pm .0054 ms$ respectively over three trials. In Fig. \ref{fig:3}b, the calibration curve of the phantom for the said experiments is shown, demonstrating a linear relationship between $\tau_{SDE}$ and $\tau_{c_{\kappa}}$ in the range of 200 $\mu s$ to 2.7 ms, with a high correlation coefficient ($R^2$) of 0.96. Data points indicated in blue were generated using Method-1 while data points indicated in red were generated using Method-2. These two methods were employed to ensure the extended dynamic range of our proposed phantom.

\subsection{Mimicking Deep tissue blood flow}
\begin{figure}[ht]
\includegraphics[height=6cm, width=13cm]{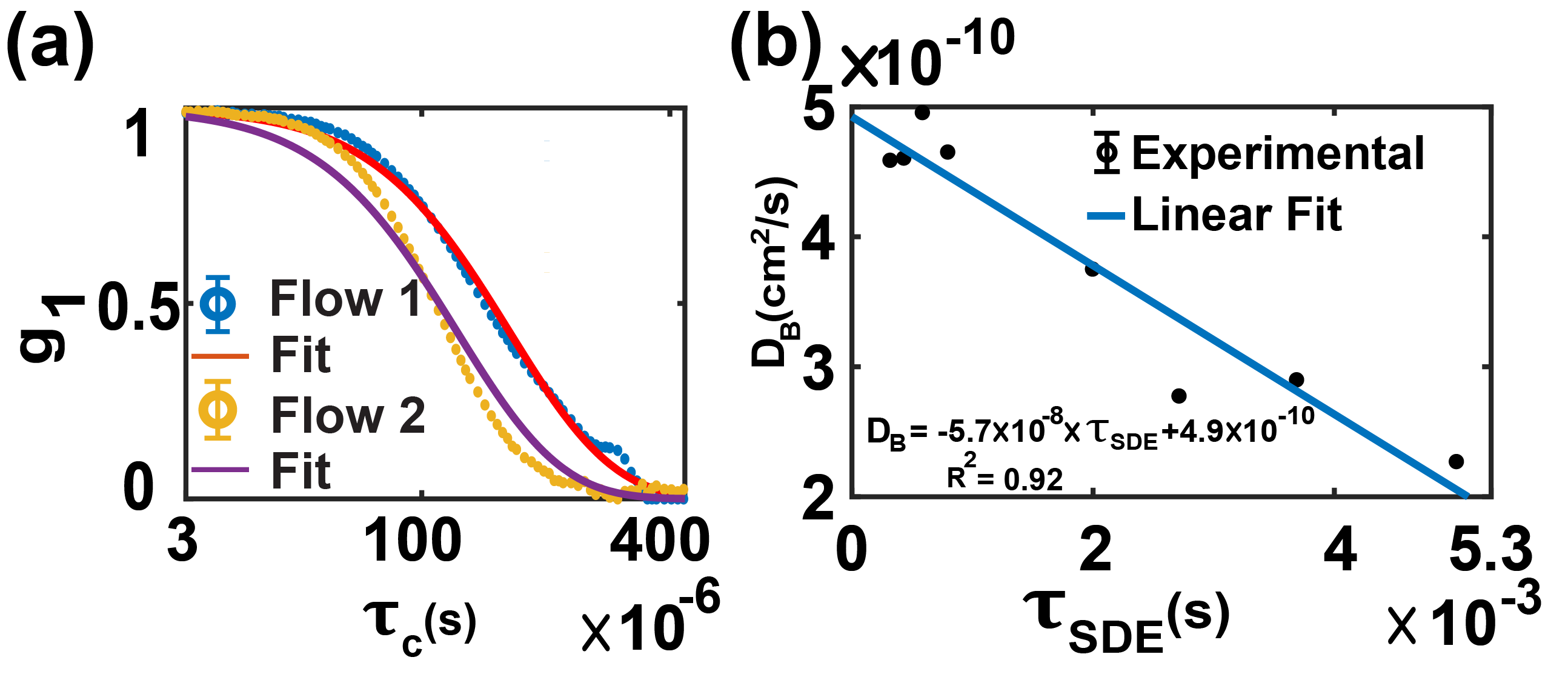}
\caption{Replicating deep tissue blood flow: (a) Mean $g_1$ values obtained from three experimental trials for two representative flows, along with their fits to semi-infinite solution of CDE. (b) Calibration curve showing a linear relationship between $D_B$ and $\tau_{SDE}$ is presented, demonstrating the potential of phantoms for deep tissue blood flow studies.}
\label{fig:4}
\end{figure}

We present the results of the proposed phantom for replicating deep tissue blood flow using DCS experiments, with a thick diffuser attached to the piezo-actuator. Fig. \ref{fig:4}(a) shows the mean $g_1$ values resulting from three trials conducted with the phantom. The representative measurements shown here were taken for two distinct values of $\tau_{SDE}$: 0.58 $ms$ and 5 $ms$, denoted as flow 1 and flow 2 respectively. The $g_1$ was fitted to the semi-infinite solution of CDE, with a reasonable agreement. Furthermore, a calibration curve, akin to the one used in MESI (as shown in Fig. \ref{fig:3}(b)) is presented in Fig. \ref{fig:4}(b).  The correlation coefficient ($R^2$) for this curve is 0.92, indicating a strong linear relationship, albeit slightly lower than the MESI curve.

\subsection{Mimicking \textit{in-vivo} human handcuff experiment }
\begin{figure}[ht]
\centering
\includegraphics[height=6cm, width=11.8cm]{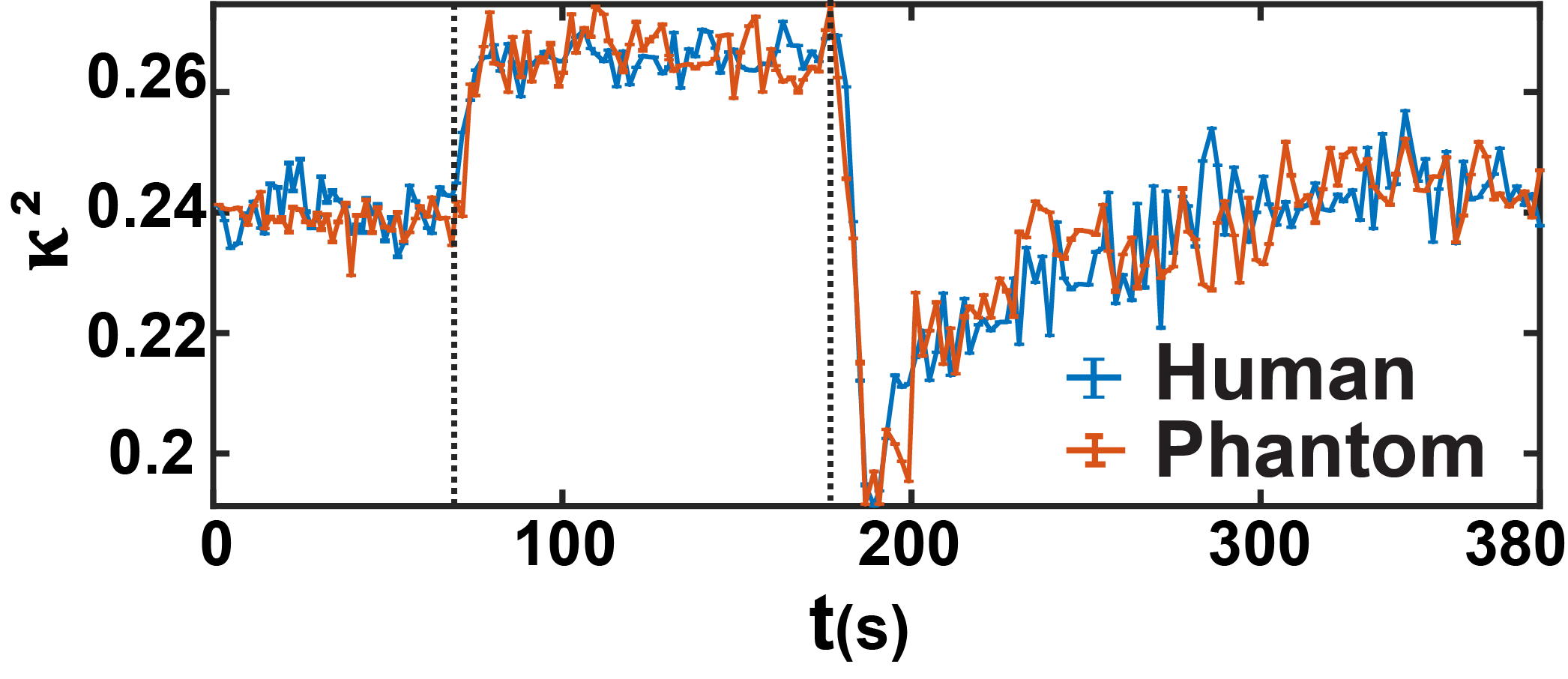}
\caption{Comparison of single-exposure speckle contrast measurements between the dynamic phantom and \textit{in-vivo} human hand cuff experiments, demonstrating the phantom's wide dynamic range and rapid tunability. The dotted lines indicate the start and end of the blood cuff in the sphygmomanometer.}
\label{Fig:5}
\end{figure}
 
Fig. \ref{Fig:5} shows the comparison of a single exposure handcuff experiment over a phase of 180 sec with three stages of baseline occlusion and post-releasing phase along with the mimicking values realized by the proposed phantom. This depicts the phantoms capability to generate such a wide range of changes in a fast and reliable manner.

\subsection{Frequency response analysis of the phantom}
\begin{figure}[ht]
\centering
\includegraphics[width=12cm,height=8cm]{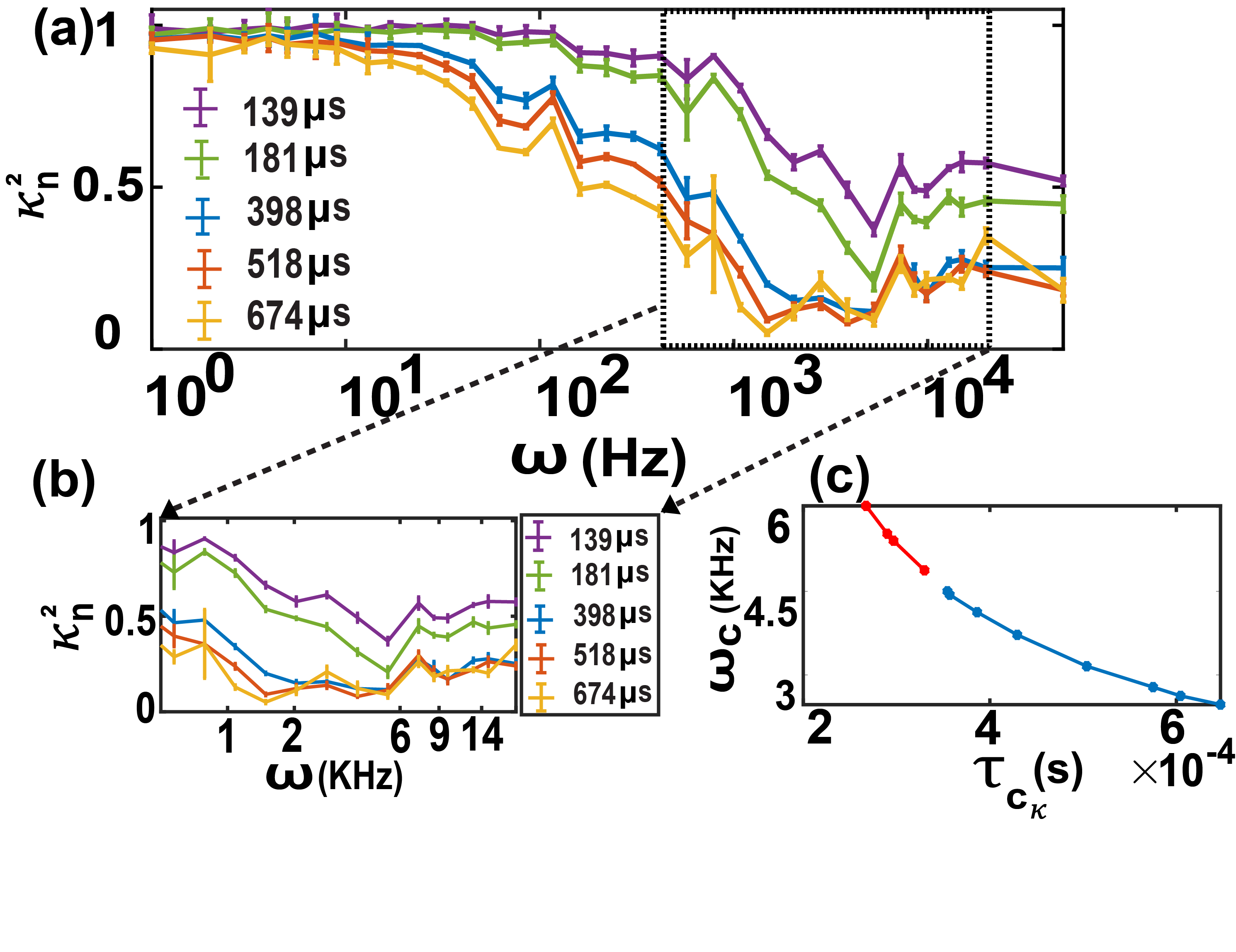}
\caption{Frequency response analysis of the phantom. (a) The relationship between input frequencies and speckle contrast at multiple exposure times, (b) a zoomed-in version revealing saturation above approximately 6 kHz, indicating the phantom's maximum operating frequency, and (c) Simulation of power spectral density.}
\label{fig:6}
\end{figure}

To determine the maximum feasible operating frequency of the phantom, we performed a frequency response analysis. For this, we applied varying sinusoidal frequencies as input signals to the proposed phantom and measured the corresponding multi-exposure speckle contrast. Fig. \ref{fig:6}(a) illustrates the outcome of this analysis, demonstrating the relationship between input frequencies and speckle contrast. A zoomed-in version of the same is shown in Fig. \ref{fig:6}(b), wherein we observed that signals approximately above 6 kHz showed signs of saturation. As a result, we concluded that the piezo-actuator along with the amplifier system we designed, has a maximum operating frequency of 6 kHz. We also conducted a power spectral density analysis using the experimentally acquired $\tau_{c_{\kappa}}$ values as shown in Fig. \ref{fig:6}(c). We simulated $g_1$ based on the experimental $\tau_{c_{\kappa}}$ values. Subsequently, we calculated its power spectrum to determine the cutoff point, denoted as $\omega_c$, which corresponds to the $-20 dB$ point from the peak amplitude. The $\omega_c$ plotted against $\tau_{c_{\kappa}}$ as in Fig. \ref{fig:6}c, confirms that the phantom's response effectively extends up to 6 kHz.

\section{Discussion}

In this paper, we have designed and developed a tunable solid phantom for laser speckle imaging to measure blood flow. To accomplish this, we employed our recently introduced SDE based algorithm \cite{murali2022laser} to generate signals $x(t)$ that control the piezo actuator. By leveraging this approach, we simulated a wide range of blood flow changes for both surface and deep tissue speckle-based imaging methods and compared them with \textit{in-vivo} human experiments.

For surface blood flow, we validated our approach by obtaining the multi-exposure speckle contrast (Fig. \ref{fig:3}a) and fitting it to an analytical solution. We also plotted the characteristic decay time obtained by the proposed phantom mimicking multiple flows (Fig. \ref{fig:3}b), which are in reasonable agreement with the desired flows, as given for simulation using SDE, with a high correlation coefficient ($R^2$) of 0.96. For the deep tissue blood flow, we obtained  the normalized field autocorrelation at a source-detector separation of 2 cm (Fig \ref{fig:4}a) and also show the inverse relationship between fitted blood flow ($D_B$) and $\tau_{SDE}$ (Fig \ref{fig:4}b). The correlation coefficient ($R^2$) for the deep tissue flow is 0.92, indicating a strong linear relationship, although it is slightly lower when compared to the surface flow case (Fig \ref{fig:3}b). This difference in correlation could be attributed to the use of a simple exponential model in SDE to generate $x(t)$. In our analysis, we acknowledge the potential for improvement by employing a complex exponential model, as proposed in Ref. \cite{murali2022laser}, for generating $x(t)$. We believe that this enhancement in modeling could lead to an increase in the correlation coefficient in deep tissue cases. Nevertheless, these results act as compelling proof of concept, demonstrating the feasibility of using dynamic phantoms to replicate flow conditions in DCS.

To demonstrate the dynamic tunability of our proposed phantom over a broad range, we obtained the speckle contrast during a human handcuff experiment (Fig. \ref{Fig:5}a). We then used the calibration curve (Fig. \ref{fig:3}b) to generate SDE signals to the piezo actuator and obtained similar speckle contrast based on the proposed phantom, which is in reasonable comparison with one another. This indicates that we can fine-tune a large dynamic range of signals based on the proposed method, limited only by the dynamic range of the piezo transducer and frame rate of the DAQ. Further based on the frequency response analysis (Fig \ref{fig:6}) of the transducer, we analyzed that the maximum operable frequency of the piezo actuator is 6 kHz, and further this range can be improved by using better piezo-actuators of higher bandwidth.

We note that the primary objective of these tunable tissue phantoms is to replicate the dynamic characteristics of the tissue, such as tissue blood flow modeled as Brownian motion, rather than mimicking optical properties like absorption and scattering coefficients. Presently, these optical properties are reproduced through the implementation of solid diffusers, attached to the piezo-actuator. Secondly, we have not explored different geometries or heterogeneous blood flow patterns. Nonetheless, we would like to note that employing a piezo-actuator array, similar to the ones demonstrated in Ref. \cite{schinaia2019novel,qiu2015piezoelectric}, holds promise for heterogeneous flow patterns by exerting independent control over each piezo actuator, which could act as individual pixels.

We emphasize the necessity of employing SDE in our work by noting that any random fluctuations alone cannot replicate blood flow without appropriately in-corporating the desired auto-correlation via SDE. For instance, when disregarding auto-correlation by setting $a=0$ and $b=1$ in the aforementioned SDE equation (1), resulting in $x(t) = dW(t)$ which is Brownian motion alone, the auto-correlation between time-instances, $t$ and $t + \tau$ yields $\sqrt{1/(1+\tau/t)}$. This relationship, as given in Chapter 4, equation 4.13 in Ref \cite{jacobs2010stochastic}, fails to exhibit the exponentially decaying auto-correlation characteristics of blood flow dynamics. 

Furthermore, it's important to acknowledge that while our study focused on employing the SDE-based algorithm for signal generation to drive the piezo actuator, alternative methods based on random signals could be explored. Approaches such as those based on COPULA \cite{duncan2008copula,frisk2024comprehensive}, Fourier transform of the phase matrix \cite{kirkpatrick2008detrimental}, correlation matrices \cite{song2016simulation}, and Karhunen–Loève expansion of electric field \cite{james2021simulation} offer promising avenues for further investigation. In comparison to the above methods, our proposed method offers a possibility to incorporate different types of auto-correlation, not restricted to simple exponential auto-correlation. Besides this, one can also incorporate different PDF by deploying Fokker-plank equations as explained in detail in our earlier work \cite{murali2022laser}. In future, the proposed phantom could be used for comparison across multiple laser speckle imaging modalities \cite{boas1995scattering,briers1982retinal,fang2024wide,kreiss2024beneath,seong2023comparison} and potentially optimizing various parameters such as exposure time and the number of frames/speckles, which are necessary for enhancing both surface and deep tissue blood flow measurement systems.

\begin{backmatter}

\bmsection{Funding}
Wadhwani Research Centre for Bioengineering (WRCB), Indian Institute of Technology Bombay; CRG–Core research grant 2023, Science Engineering Research Board - Department of Science and Technology, Ministry of Science and Technology, India.

\bmsection{Disclosures}

\noindent The authors declare no conflicts of interest. Portions of this work were presented at the European Conferences on Biomedical Optics (ECBO), 2023, in Ref\cite{sarkar2023development}.

\bmsection{Data availability} Data underlying the results presented in this paper are not publicly available at this time but may be obtained from the authors upon reasonable request.

\end{backmatter}

\bibliography{Paper.bib}

\end{document}